\documentclass[american,english,twocolumn,amsmath,amssymb,prl,showpacs,superscriptaddress]{revtex4}
\usepackage[latin1]{inputenc}
\usepackage{color}
\usepackage{graphicx}

\makeatletter
\usepackage{dcolumn}
\usepackage{bm}
\usepackage{rotating}
\usepackage{psfrag}       
\def\kbar{{\mathchar'26\mkern-9mu k}}

\makeatother

\usepackage{babel}

\begin{document}

\title{Experimental observation of the Anderson transition with atomic matter
waves}

\author{Julien Chab{\'e}}

\affiliation{Laboratoire de Physique des Lasers, Atomes et Mol{\'e}cules, Universit{\'e}
des Sciences et Technologies de Lille, CNRS; CERLA; F-59655 Villeneuve
d'Ascq Cedex, France}

\selectlanguage{american}

\homepage{www.phlam.univ-lille1.fr/atfr/cq}

\author{Gabriel Lemari{\'e}}

\author{Beno{\^i}t Gr{\'e}maud}

\author{Dominique Delande}

\affiliation{Laboratoire Kastler Brossel, Universit{\'e} Pierre et Marie Curie, 4
Place Jussieu, F-75005 Paris, France}

\author{Pascal Szriftgiser}

\author{Jean Claude Garreau }

\affiliation{Laboratoire de Physique des Lasers, Atomes et Mol{\'e}cules, Universit{\'e}
des Sciences et Technologies de Lille, CNRS; CERLA; F-59655 Villeneuve
d'Ascq Cedex, France}

\homepage{www.phlam.univ-lille1.fr/atfr/cq}

\date{\today}

\begin{abstract}
We realize experimentally an atom-optics quantum chaotic system, the
quasiperiodic kicked rotor, which is equivalent to a 3D disordered
system, that allow us to demonstrate the Anderson metal-insulator
transition. Sensitive measurements of the atomic wavefunction dynamics
and the use of finite-size scaling techniques make it possible to
extract both the critical parameters and the critical exponent of
the transition, which is in good agreement with the value obtained
in numerical simulations of the 3D Anderson model.
\end{abstract}

\pacs{03.75.-b, 72.15.Rn, 05.45.Mt, 64.70.Tg}

\maketitle
The metal-insulator Anderson transition plays a central role in the
study of quantum disordered systems. An insulator is associated with
localized states of the system, while a metal generally displays diffusive
transport associated with delocalized states. The Anderson model \citep{Anderson:LocAnderson:PR58}
describes such a metal-insulator transition, due
to quantum interference effects driven by the amount of disorder
in the system. Starting from the ``tight-binding'' description
of an electron in a crystal lattice, Anderson postulated in 1958 that
the dominant effect of impurities in the lattice is to randomize the
diagonal, on-site, term of the Hamiltonian, and showed that this generally
leads to a localization of the wavefunction, in sharp contrast with
the Bloch-wave solution for a perfect crystal. This model has progressively
been extended from its original solid-state physics 
scope \citep{Anderson:LocAnderson:PR58,Altshuler:MetalInsulator:ANP06,Kramer:Localization:RPP93,Thouless:AndLoc:PREP74}
to a whole class of systems in which waves propagate in a disordered
medium, as for example quantum-chaotic systems \citep{Casati:LocDynFirst:LNP79,Fishman:LocDynAnderson:PRA84}
and electromagnetic radiation \citep{Maret:AndersonTransLight:PRL06,Segev:LocAnderson2DLight:N07,Dembowski:AndersonMicrocavity:PRE99}.
However mathematically simple, the model predicts a wealth of interesting
phenomena. In 1D, the wavefunction is always localized as recently
observed in experiments using atomic matter waves in a disordered
optical potential \citep{Bouyer:AndersonBEC:N08,Inguscio:AndersonBEC:N08};
in 3D it predicts a phase transition between a localized (insulator)
and a delocalized (metal) phase at a well defined mobility edge, the
density of impurities or the energy being the control parameter.

Despite the wide interest on the Anderson transition, few experimental
results are available. In a crystal, it is very difficult to obtain
the conditions for a clean observation of the Anderson localization.
Firstly, one has no direct access to the electronic wavefunction and
must rely on modifications of bulk properties like conductivity 
\citep{Altshuler:MetalInsulator:ANP06,MacKinnon:CriticalExp:JPC94}.
Secondly, it is difficult to reduce decoherence sources to a low enough
level. We thus engineered a matter-wave system that is described by
an Anderson-like model, which allows us to probe the physics of disordered
systems in much better conditions than in condensed matter physics
\citep{Lewenstein:UltracoldSolidState:ADVP07}, namely: Almost no
interaction between particles, weak absorption in the medium, no coupling
with a thermal reservoir which could destroy localization and possibility
of measuring the final quantum state of the system after a given interaction
time. The system, the quasi-periodic kicked rotor 
\citep{Casati:LocDynFirst:LNP79,Fishman:LocDynAnderson:PRA84,Raizen:QKRFirst:PRL95,Casati:IncommFreqsQKR:PRL89},
consists of cold cesium atoms exposed to a pulsed, off-resonant, laser
standing wave. The dynamics is thus effectively one-dimensional
along the axis of the laser beam, as transverse directions are uncoupled.
The atoms interact periodically with the spatially sinusoidal potential
whose amplitude is modulated at (incommensurable) frequencies $\omega_{2}$
and $\omega_{3}$, and the corresponding Hamiltonian is: 
\begin{equation}
H=\frac{p^{2}}{2}+K\cos x\left[1+\varepsilon\cos\left(\omega_{2}t\right)\cos\left(\omega_{3}t\right)\right]\sum_{n=0}^{N-1}\delta(t-n)\;,
\label{H}
\end{equation}
where $x$ is the particle position (along the laser axis), $p$ is
its momentum, and $K$ is the pulse intensity. The pulses are short
enough (compared to the center of mass dynamics) to be consider as
instantaneous kicks. We have chosen normalized variables
such that $x$ is measured in units of the spatial period of the potential
divided by $2\pi$, the particle's mass is unity and time is measured
in units of pulse period $T_{1}$.

By taking $\varepsilon=0$ in Eq.~(\ref{H}) one obtains the standard
(strictly time-periodic) kicked rotor, a system known to display the
phenomenon of \textit{dynamical localization} \citep{Casati:LocDynFirst:LNP79},
which manifests itself by an exponential localization of the wavefunction
in \textit{momentum space}. Dynamical localization exists only when
the classical dynamics of the kicked rotor is a chaotic diffusion.
In the quantum case, diffusion is inhibited after some localization
time by quantum interferences. Dynamical localization has been shown
to be a direct analogue of Anderson localization in one dimension
\citep{Fishman:LocDynAnderson:PRA84}, with the following correspondences: Localization takes place
in real space for the Anderson model and in momentum space for the
kicked rotor; the ``stochasticity parameter'' $K$ {[}see Eq.~(\ref{H})]
corresponds to the ratio of hopping to diagonal energy in the Anderson
model. The strictly random disorder in the Anderson model correspond
to a pseudo-random potential in the quasi-periodic kicked rotor. Thus,
the experimental observation of dynamical localization in the kicked
rotor \citep{Raizen:QKRFirst:PRL95} is actually the first observation
of Anderson 1D localization with atomic matter waves.

\begin{figure}
\begin{centering}
\includegraphics[width=7cm]{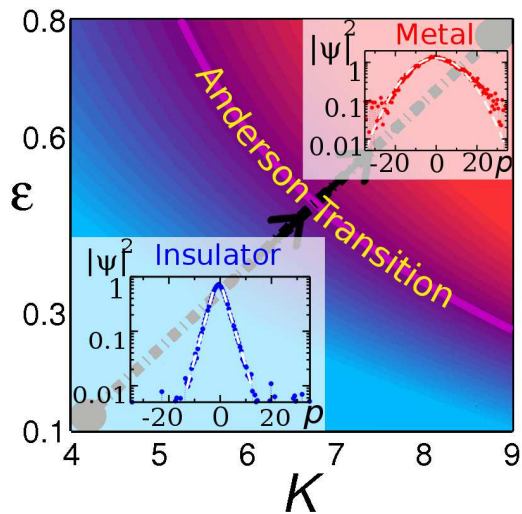} 
\par\end{centering}

\caption{
\label{fig:K-Epsilon} Phase diagram of the quasiperiodic kicked rotor,
from numerical simulations. The localized (insulator) region is shown
in blue, the diffusive (metallic) region is shown in red. The experimental
parameters are swept along the diagonal dash-dotted line. The insets
show the \textit{experimentally observed} momentum distributions,
localized in the insulator region and Gaussian in the diffusive (metallic)
region. }

\end{figure}

In order to observe the Anderson transition, one must generalize the
kicked rotor to obtain a system analogous to the 3D Anderson
model. This can be done by introducing further temporal dependencies
that make the Hamiltonian quasi-periodic in time, that is to take
$\varepsilon\neq0$ in Eq.~(\ref{H}) with $\omega_{2}$ and $\omega_{3}$
incommensurate numbers. The resulting system has been shown to be
substantially equivalent to the 3D Anderson model \citep{Casati:IncommFreqsQKR:PRL89},
the additional spatial coordinates in the Anderson model corresponding
to the additional time dependencies in the quasi-periodic kicked rotor.
The parameters controlling the dynamics are $\omega_{2}$ and $\omega_{3}$
and, more importantly, the kick strength $K$ and the modulation amplitude
$\varepsilon$. In the $(K,\varepsilon)$ plane, the $\varepsilon=0$
line corresponds to dynamically localized dynamics (the periodic kicked
rotor equivalent to the 1D Anderson model) and the Anderson transition
takes place along a critical line in the plane $\varepsilon>0$, as
shown in Fig.~\ref{fig:K-Epsilon}.

Our atom-optics realization of the kicked rotor has been described
in detail elsewhere \citep{AP:ChaosQTransp:CNSNS:2003,AP:RamanSpectro:PRA01,AP:DiodeMod:EPJD99}.
Basically, we cool cesium atoms in a standard magneto-optical trap,
and, after a Sisyphus-molasses phase, we obtain a cloud of $10^{7}$
atoms at a temperature of $3.2\,\mathrm{\mu K}$. This prepares a
sample of atoms in a thermal state whose momentum distribution is
much narrower than the expected localization length. The atoms then
interact with the optical potential generated by a horizontal standing
wave. An acousto-optical modulator driven by an arbitrary-form synthesizer
modulates the amplitude of the optical potential. One generates in
this way $0.9\,\mathrm{\mu s}$-long pulses at $T_{1}=27.778\,\mathrm{\mu s}$
(corresponding to an effective Planck constant $\kbar=2.89$), to
which is superimposed a modulation of the form Eq.~(\ref{H}), with
$\omega_{2}/2\pi=\sqrt{5}$ and $\omega_{3}/2\pi=\sqrt{13}$. The
standing wave, of typical power $160\,\mathrm{mW}$, is far off-resonant
($7.3\,\mathrm{GHz}$ to red of the atomic transition, corresponding
to $1.4\times10^{3}$ natural widths), in order to reduce spontaneous
emission. Stimulated Raman transitions allow sensitive measurements
of the atomic momentum distribution \citep{AP:RamanSpectro:PRA01}.

The experimental values of the parameters are chosen according to
the following considerations. Firstly, in order to prevent classical (KAM
barrier) localization effects one must have $K>2$ \citep{note:chaotic}.
For all the data presented in this paper, we numerically checked that
the classical dynamics is fully diffusive, which excludes that the
observed transition could have a classical origin. Secondly, in order
to confine the transition to a relatively narrow range of parameters
one must cross the critical curve (Fig.~\ref{fig:K-Epsilon}) ``at
a right angle''; we thus vary simultaneously the $K$ and $\varepsilon$
along a line going from $K=4$, $\varepsilon=0.1$ in the localized
region to $K=9$, $\varepsilon=0.8$ in the diffusive region; the
critical line is then crossed at $K=K_{c}=6.6.$ Thirdly, short enough
pulses must be used, so that they can be considered as delta pulses.
Numerical simulations taking into account the finite pulse duration
($0.9\,\mathrm{\mu s}$) show that less than 1\% of the atoms are
sensitive to the duration of the pulses. Finally, decoherence processes
must be kept as small as possible. We have identified two dominant
decoherence sources in our experiment: i) Spontaneous emission, which
is not included in the Hamiltonian Eq.~(\ref{H}) and ii) the deviation
of the standing wave from strict horizontality, which mixes quasimomentum
classes and produces a stray momentum diffusion. The fact that numerical
simulations of the ``pure'' quasi-periodic kicked rotor shown
hereafter agree very well with the experimental results (in particular
for the position of the Anderson transition and for the critical exponent)
proves that spurious effects are well under control. Moreover, we
have checked that inclusion of these effects in the numerical calculations
only leads to small changes for $t\leq150$ kicks.

\begin{figure}
\begin{centering}
\psfrag{t (number of kicks)}{\begin{large}$t$ (number of kicks)                                \end{large}}
\psfrag{Pi}{\begin{large}$\dfrac{1}{{\Pi_0}^2}$                                 \end{large}}
\includegraphics[width=8cm]{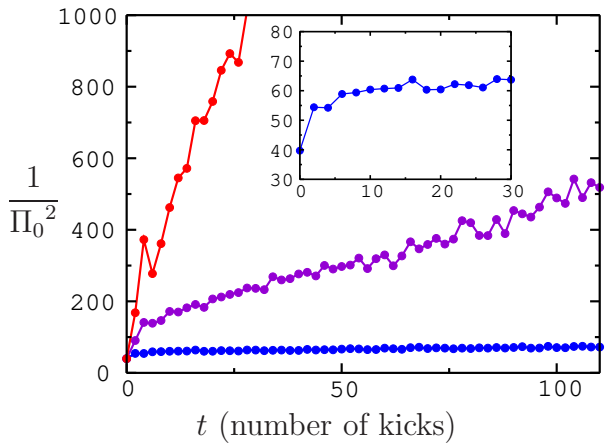} 
\par\end{centering}

\caption{
\label{fig:KineticEnergy} Temporal dynamics of the quasi-periodically
kicked rotor. We measure the population $\Pi_{0}(t)$ of the zero-momentum
class as a function of time and plotted the quantity $\Pi_{0}^{-2}(t)$
(proportional to $\langle p^{2}\rangle(t)$). Clearly, it tends to
a constant in the localized regime (blue curve) and increases linearly
with time in the diffusive regime (red curve) . Close to the critical
point (purple curve), it displays anomalous diffusion $\Pi_{0}^{-2}(t)\sim t^{2/3}$,
as expected from theoretical arguments \citep{Ohtsuki:AndersonTrans:JPSJ97}.}

\end{figure}

In order to observe the Anderson transition we apply a sequence of
kicks to the atomic cloud and measure its dynamics. In the localized
regime, the evolution of its momentum distribution is ``frozen''
after the localization time ($\sim$12 kicks) into an exponential curve
$\exp\left(-|p|/p_{\text{loc}}\right)$ (where $p_{\text{loc}}$ is
the \textit{localization length}). In the diffusive regime, the initial
Gaussian shape is preserved and the distribution gets broader as kicks
are applied, corresponding to a linear increase of the average kinetic
energy. The insets in Fig.~\ref{fig:K-Epsilon} show the experimentally
observed momentum distributions, an exponentially localized distribution
for $K<K_{c}$ (blue curve), characteristic of dynamical localization,
and a broad, Gaussian-shaped distribution for $K>K_{c}$ (red curve),
characteristic of the diffusive regime. Instead of measuring the full
momentum distribution, it is sufficient, and much easier, to measure
the population $\Pi_{0}(t)$ of the zero velocity class \citep{note:Raman},
as $\Pi_{0}^{-2}(t)$ is proportional to $\langle p^{2}\rangle(t)$
(by virtue of the constancy of the total number of atoms). We performed
several experimental runs corresponding to points in the $(K,\epsilon)$ plane.
In each run the value of $\Pi_{0}(t)$ was recorded as the kicks were
applied. We also recorded the background signal obtained by not applying
the Raman detection sequence, and the total number of atoms in the
cold-atom cloud. These signals are used to correct the experimental
data from background signals and long-term drifts of the cloud population.
Figure \ref{fig:KineticEnergy} shows the experimentally observed
$\Pi_{0}^{-2}(t)$ and clearly shows the transition from the localized
to the diffusive regime, with an intermediate regime of anomalous
diffusion.

When one approaches the critical point from the insulator side, the
localization length diverges, whereas, on the metallic side, the diffusion
constant vanishes. However, a strict divergence can be observed only
in macroscopically large samples; in small samples the divergence
is smoothed. This fact plagued the numerical studies of the solid-state
Anderson transition, as only a finite (small) lattice can be dealt
with in a computer. In our system, a singular behavior would show
up only for prohibitively large numbers of kicks, which are, in practice,
limited to 150. To overcome this limitation, a technique named ``finite-size
scaling'' \citep{FisherBarber:FiniteSizeScal:PRL72,Slevin:ScalingAnderson:PRL99,MacKinnon:CriticalExp:JPC94}
was introduced, whose basic idea is to infer the scaling law allowing
proper extrapolation of the measured localization length to an infinite
sample. We will next show that our data obey the scaling laws predicted
by renormalization theory, and this will allow us to extract the critical
exponent associated with the singular behavior at the transition. 

\begin{figure*}
\begin{centering}
\includegraphics[width=4.2cm]{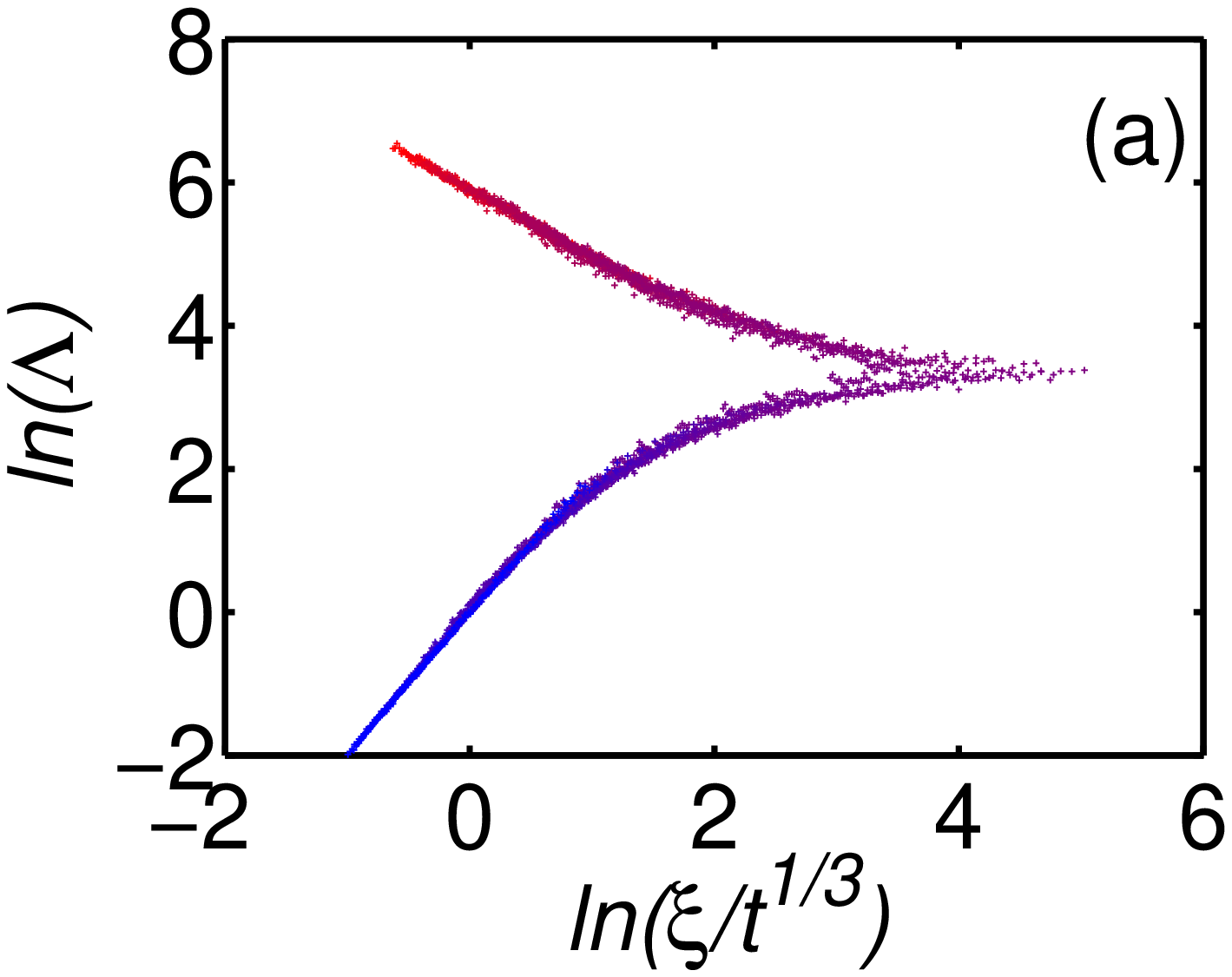} \includegraphics[width=4.2cm]{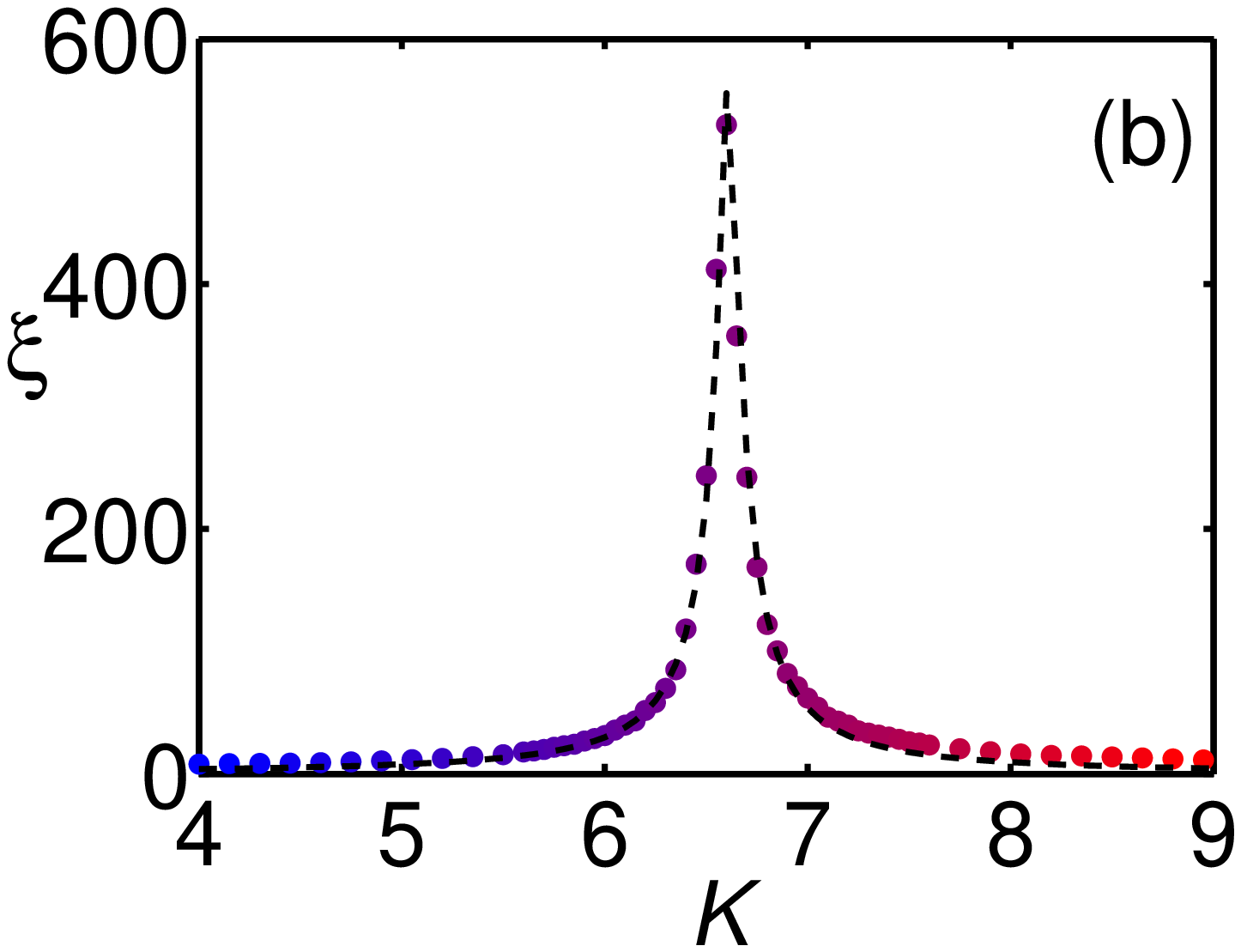}
\includegraphics[width=4.2cm]{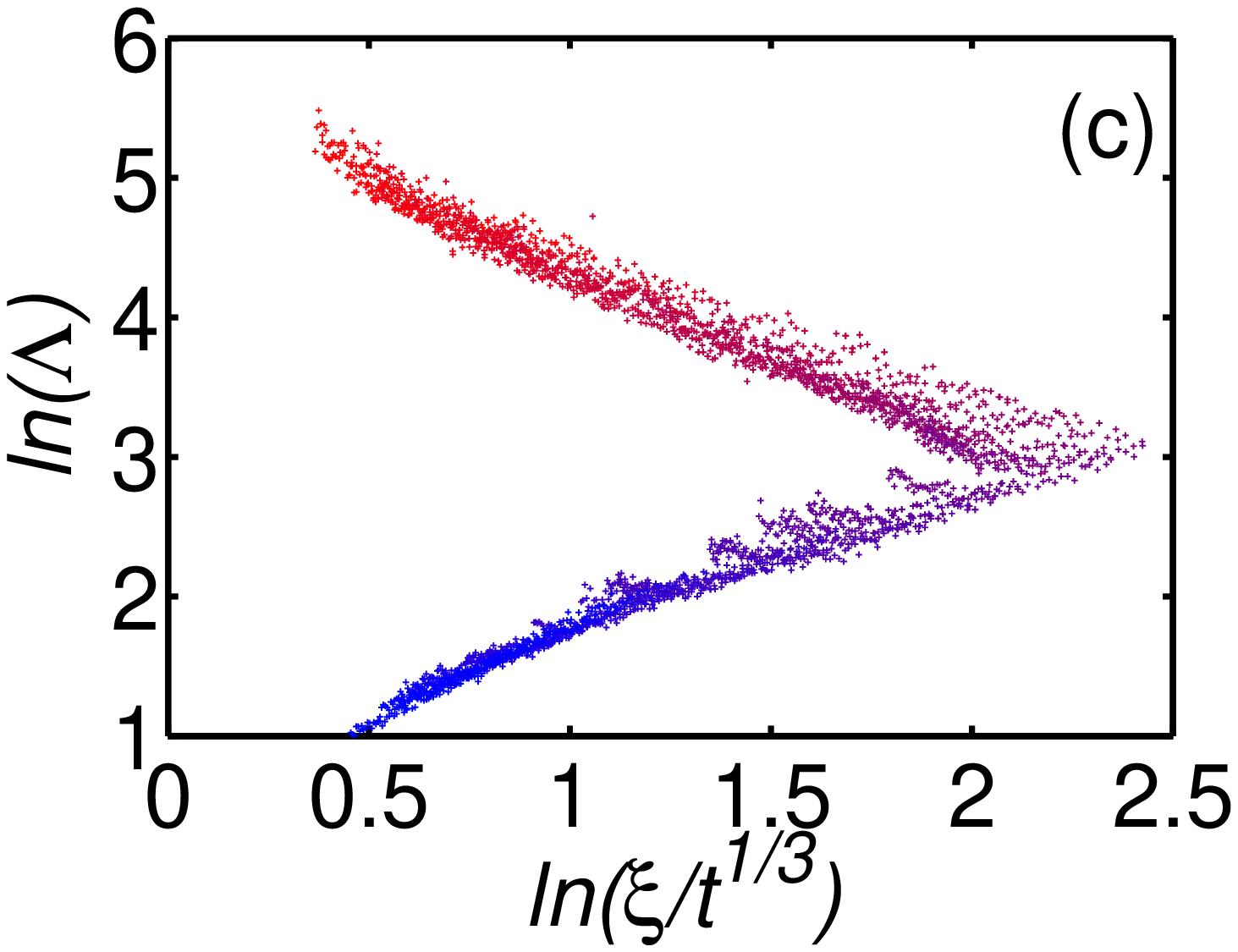} \includegraphics[width=4.2cm]{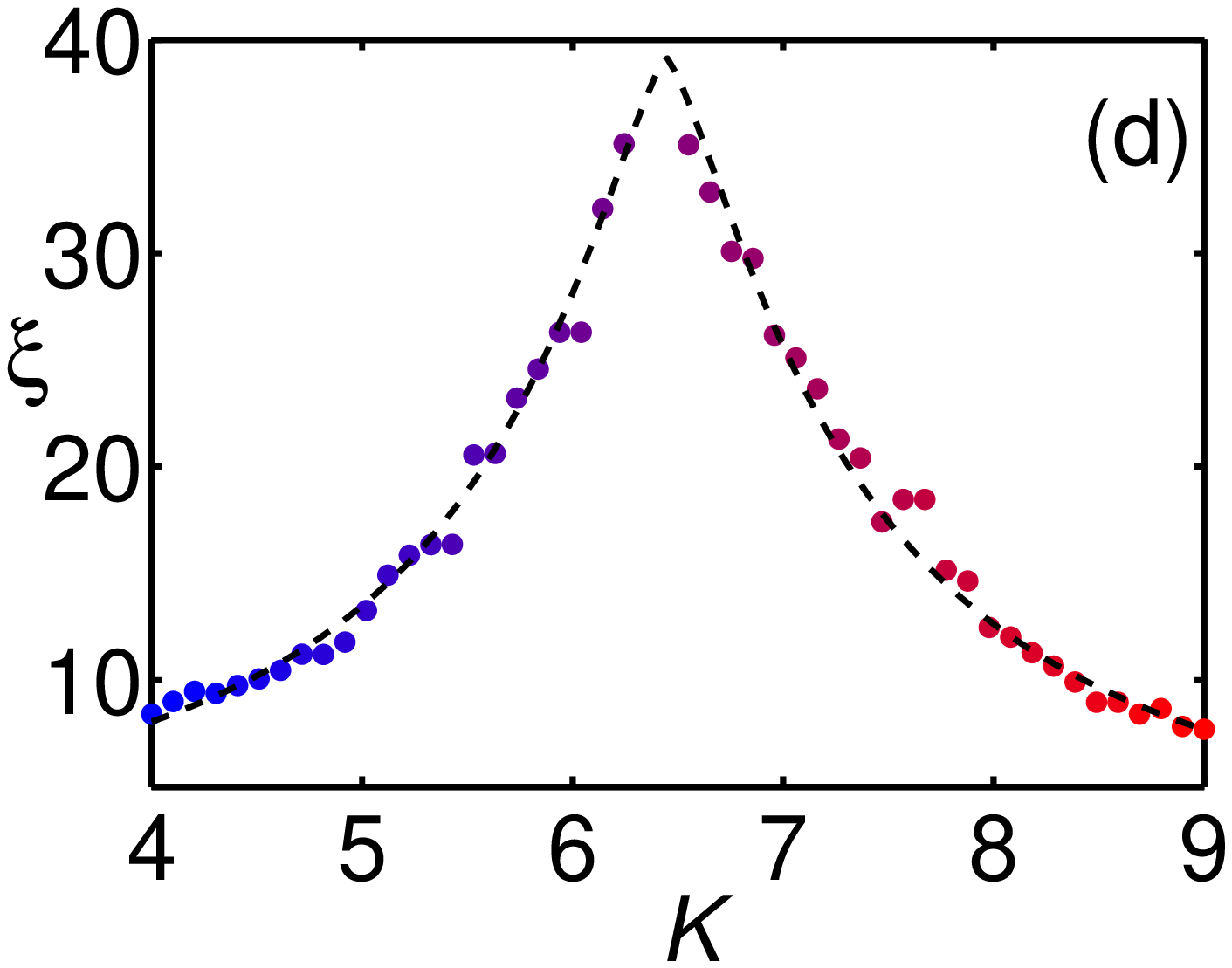} 
\par\end{centering}

\caption{
\label{fig:FiniteSizeScaling} Finite size scaling applied to the
results of numerical simulations {[}(a) and (b)] and to the experimental
results {[}(c) and (d)]. The rescaled quantity $\Lambda=\Pi_{0}^{-2}(t)t^{-2/3}$
is plotted as a function of time (from $30$ to $10^{4}$ kicks (simulation)
and from $30$ to $150$ kicks (experiment), for various values of $K$.
Finite-size scaling of the raw data makes it possible to determine
both the scaling function $f$, shown in (a) and (c), and the scaling
parameter $\xi(K)$, represented in (b) and (d). Close to $K_{c}$,
the behavior of $\xi$ in (b) and (d) is well fitted by Eq.~(\ref{eq:xi})
(dashed lines), giving $K_{c}=6.6$ (simulation) and $K_{c}=6.4$
(experiment). The critical exponent is $\nu=1.60\pm0.05$ (simulation)
and $\nu=1.4\pm0.3$ (experiment).}

\end{figure*}

We have adapted the standard finite-size scaling approach used in
numerical studies of the Anderson transition \citep{Kramer:Localization:RPP93}
assuming that, for finite interaction time, the quantity $\Lambda(t)=\Pi_{0}^{-2}(t)t^{-2/3}$
is an arbitrary function $f\left(\xi t^{-1/3}\right)$ 
\citep{Wegner:ScalingMobilityEdge:ZFP76,Ohtsuki:AndersonTrans:JPSJ97,Stauffer:Percolation:94},
with a scaling parameter $\xi$ which depends \textit{only} on $K$.
Using the results obtained for various values of $t$ and $K$ one
can reconstruct both the function $f$ and the scaling parameter $\xi(K)$
(no assumption on the form of $f$ is made). The result is shown in
Fig.~\ref{fig:FiniteSizeScaling}(a) and (b) for numerical simulations
and in Fig.~\ref{fig:FiniteSizeScaling}(c) and (d) for the experimental
results. In both cases, the scaling hypothesis is justified by the
fact that all points in Fig.~\ref{fig:FiniteSizeScaling}(a) and \ref{fig:FiniteSizeScaling}(c)
lie on a single curve (within the experimental error). The remarkable
feature is the existence of two branches: the upper one corresponds
to diffusive motion while the lower one corresponds to the localized
regime, the critical point being at the tip joining the two branches.
The scaling parameter $\xi(K)$ is plotted in Figs. \ref{fig:FiniteSizeScaling}(b)
and \ref{fig:FiniteSizeScaling}(d): It represents the localization
length in the localized regime and scales as the inverse of the diffusion
constant in the diffusive regime. Clearly, it increases rapidly in
the vicinity of the critical value $K_{c}$, on both sides of the
transition, and it is found to behave as $\xi\sim\vert K-K_{c}\vert^{-\nu}$
when $K\rightarrow K_{c}$. This divergence at the critical point
is a key property of the Anderson phase transition. However, phase-breaking
mechanisms induced by decoherence processes smooth the algebraic divergence.
We model such effects by introducing a small cutoff of the divergence
due to the residual decoherence: 
\begin{equation}
\frac{1}{\xi(K)}=\alpha\vert K-K_{c}\vert^{\nu}+\beta\;.
\label{eq:xi}
\end{equation}
The experimental observations and the numerical data have been fitted
with this formula {[}dashed curves in Figs. \ref{fig:FiniteSizeScaling}(b)
and \ref{fig:FiniteSizeScaling}(d)]. We found $K_{c}=6.4$ (very
close to the value $K_{c}=6.6$ obtained from the numerical simulation),
and a critical exponent $\nu=1.4\pm0.3$ \citep{note:CriticalExp},
which is consistent with the numerical value within the error bars.
The good agreement between the numerical simulations and the experimental
results proves that spurious effects (such as decoherence) are under
control. We emphasize that there are no adjustable parameters in our
procedure, all parameters are determined using the atoms themselves
as probes. Once the existence of the scaling law established,
it is more convenient to use a \textit{global} analysis of the numerical data at various
values of $K$ (see refs. \citet{Slevin:ScalingAnderson:PRL99,MacKinnon:CriticalExp:JPC94}).
This, together with the fact that numerical simulations are not limited
to 150 kicks and can be ran up to several thousands kicks, make it possible
to obtain a more precise numerical value for the critical exponent
$\nu=1.60\pm0.05$. Note, finally, that although the data displayed
here concerns a particular set of parameters we also verified experimentally
the presence of the transition for other parameters.

In conclusion, we have presented the first experimental evidence of
the Anderson transition with atomic matter waves. The transition is
characterized by a well defined critical point, a divergence of the
localization length below the critical point (in the localized regime)
and a vanishing of the diffusion constant above the critical point
(in the diffusive regime). We have determined the scaling laws and
the critical exponent $\nu$  of the Anderson transition, which is significantly
larger than unity and very close to the value obtained in recent numerical
experiments \citep{Slevin:ScalingAnderson:PRL99,MacKinnon:CriticalExp:JPC94}
on the Anderson model, enforcing the assumption \citep{Casati:IncommFreqsQKR:PRL89}
that the two systems are substantially equivalent. Whether this exponent
is universal (i.e. independent of the microscopic details) or not
remains to be studied. A very interesting point is that our Anderson-equivalent
quasiperiodic kicked rotor can be easily generalized to higher dimensions
simply by adding new incommensurable frequency, which opens perspectives
for fascinating studies of the dependence of the critical exponent
on the dimension of the underlying Anderson model.

\begin{acknowledgments}
The authors acknowledge D. Shepelyansky for bringing ref. \citep{Casati:IncommFreqsQKR:PRL89}
to their attention, and for fruitful discussions. They also acknowledge
G. Beck for his help with the experiment.
\end{acknowledgments}

%\bibliographystyle{apsrev}
%\bibliography{ArtDataBase,notes}

\begin{thebibliography}{26}
\expandafter\ifx\csname natexlab\endcsname\relax\def\natexlab#1{#1}\fi
\expandafter\ifx\csname bibnamefont\endcsname\relax
  \def\bibnamefont#1{#1}\fi
\expandafter\ifx\csname bibfnamefont\endcsname\relax
  \def\bibfnamefont#1{#1}\fi
\expandafter\ifx\csname citenamefont\endcsname\relax
  \def\citenamefont#1{#1}\fi
\expandafter\ifx\csname url\endcsname\relax
  \def\url#1{\texttt{#1}}\fi
\expandafter\ifx\csname urlprefix\endcsname\relax\def\urlprefix{URL }\fi
\providecommand{\bibinfo}[2]{#2}
\providecommand{\eprint}[2][]{\url{#2}}

\bibitem[{\citenamefont{Anderson}(1958)}]{Anderson:LocAnderson:PR58}
\bibinfo{author}{\bibfnamefont{P.~W.} \bibnamefont{Anderson}},
  \bibinfo{journal}{Phys. Rev.} \textbf{\bibinfo{volume}{109}},
  \bibinfo{pages}{1492} (\bibinfo{year}{1958}).

\bibitem[{\citenamefont{Basko et~al.}(2006)\citenamefont{Basko, Aleiner, and
  Altshuler}}]{Altshuler:MetalInsulator:ANP06}
\bibinfo{author}{\bibfnamefont{D.~M.} \bibnamefont{Basko}},
  \bibinfo{author}{\bibfnamefont{I.~L.} \bibnamefont{Aleiner}},
  \bibnamefont{and} \bibinfo{author}{\bibfnamefont{B.~L.}
  \bibnamefont{Altshuler}}, \bibinfo{journal}{Ann. Phys.}
  \textbf{\bibinfo{volume}{321}}, \bibinfo{pages}{1126} (\bibinfo{year}{2006}).

\bibitem[{\citenamefont{Kramer and
  Mackinnon}(1993)}]{Kramer:Localization:RPP93}
\bibinfo{author}{\bibfnamefont{B.}~\bibnamefont{Kramer}} \bibnamefont{and}
  \bibinfo{author}{\bibfnamefont{A.}~\bibnamefont{Mackinnon}},
  \bibinfo{journal}{Rep. Prog. Phys.} \textbf{\bibinfo{volume}{56}},
  \bibinfo{pages}{1469} (\bibinfo{year}{1993}).

\bibitem[{\citenamefont{Thouless}(1974)}]{Thouless:AndLoc:PREP74}
\bibinfo{author}{\bibfnamefont{D.~J.} \bibnamefont{Thouless}},
  \bibinfo{journal}{Phys. Rep.} \textbf{\bibinfo{volume}{13}},
  \bibinfo{pages}{93} (\bibinfo{year}{1974}).

\bibitem[{\citenamefont{Casati et~al.}(1979)\citenamefont{Casati, Chirikov,
  Ford, and Izrailev}}]{Casati:LocDynFirst:LNP79}
\bibinfo{author}{\bibfnamefont{G.}~\bibnamefont{Casati}},
  \bibinfo{author}{\bibfnamefont{B.~V.} \bibnamefont{Chirikov}},
  \bibinfo{author}{\bibfnamefont{J.}~\bibnamefont{Ford}}, \bibnamefont{and}
  \bibinfo{author}{\bibfnamefont{F.~M.} \bibnamefont{Izrailev}},
  \textit{\bibinfo{title}{{Stochastic behavior of a quantum pendulum under
  periodic perturbation}}} (\bibinfo{publisher}{{Springer-Verlag}},
  \bibinfo{address}{{Berlin, Germany}}, \bibinfo{year}{1979}),
  vol.~\bibinfo{volume}{93}, pp. \bibinfo{pages}{334--352}.

\bibitem[{\citenamefont{Grempel et~al.}(1984)\citenamefont{Grempel, Prange, and
  Fishman}}]{Fishman:LocDynAnderson:PRA84}
\bibinfo{author}{\bibfnamefont{D.~R.} \bibnamefont{Grempel}},
  \bibinfo{author}{\bibfnamefont{R.~E.} \bibnamefont{Prange}},
  \bibnamefont{and} \bibinfo{author}{\bibfnamefont{S.}~\bibnamefont{Fishman}},
  \bibinfo{journal}{Phys. Rev. A} \textbf{\bibinfo{volume}{29}},
  \bibinfo{pages}{1639} (\bibinfo{year}{1984}).

\bibitem[{\citenamefont{St{\"o}rzer et~al.}(2006)\citenamefont{St{\"o}rzer,
  Gross, Aegerter, and Maret}}]{Maret:AndersonTransLight:PRL06}
\bibinfo{author}{\bibfnamefont{M.}~\bibnamefont{St{\"o}rzer}},
  \bibinfo{author}{\bibfnamefont{P.}~\bibnamefont{Gross}},
  \bibinfo{author}{\bibfnamefont{C.~M.} \bibnamefont{Aegerter}},
  \bibnamefont{and} \bibinfo{author}{\bibfnamefont{G.}~\bibnamefont{Maret}},
  \bibinfo{journal}{Phys. Rev. Lett.} \textbf{\bibinfo{volume}{96}},
  \bibinfo{pages}{063904} (\bibinfo{year}{2006}).

\bibitem[{\citenamefont{Schwartz et~al.}(2007)\citenamefont{Schwartz, Bartal,
  Fishman, and Segev}}]{Segev:LocAnderson2DLight:N07}
\bibinfo{author}{\bibfnamefont{T.}~\bibnamefont{Schwartz}},
  \bibinfo{author}{\bibfnamefont{G.}~\bibnamefont{Bartal}},
  \bibinfo{author}{\bibfnamefont{S.}~\bibnamefont{Fishman}}, \bibnamefont{and}
  \bibinfo{author}{\bibfnamefont{B.}~\bibnamefont{Segev}},
  \bibinfo{journal}{Nature (London)} \textbf{\bibinfo{volume}{446}},
  \bibinfo{pages}{52} (\bibinfo{year}{2007}).

\bibitem[{\citenamefont{Dembowski et~al.}(1999)\citenamefont{Dembowski,
  Gr{\"a}f, Hofferbert, Rehfeld, Richter, and
  Weiland}}]{Dembowski:AndersonMicrocavity:PRE99}
\bibinfo{author}{\bibfnamefont{C.}~\bibnamefont{Dembowski}},
  \bibinfo{author}{\bibfnamefont{H.~D.} \bibnamefont{Gr{\"a}f}},
  \bibinfo{author}{\bibfnamefont{R.}~\bibnamefont{Hofferbert}},
  \bibinfo{author}{\bibfnamefont{H.}~\bibnamefont{Rehfeld}},
  \bibinfo{author}{\bibfnamefont{A.}~\bibnamefont{Richter}}, \bibnamefont{and}
  \bibinfo{author}{\bibfnamefont{T.}~\bibnamefont{Weiland}},
  \bibinfo{journal}{Phys. Rev. E} \textbf{\bibinfo{volume}{60}},
  \bibinfo{pages}{3942} (\bibinfo{year}{1999}).

\bibitem[{\citenamefont{Billy et~al.}(2008)\citenamefont{Billy, Josse, Zuo,
  Bernard, Hambrecht, Lugan, Cl{\'e}ment, Sanchez-Palencia, Bouyer, and
  Aspect}}]{Bouyer:AndersonBEC:N08}
\bibinfo{author}{\bibfnamefont{J.}~\bibnamefont{Billy}},
  \bibinfo{author}{\bibfnamefont{V.}~\bibnamefont{Josse}},
  \bibinfo{author}{\bibfnamefont{Z.}~\bibnamefont{Zuo}},
  \bibinfo{author}{\bibfnamefont{A.}~\bibnamefont{Bernard}},
  \bibinfo{author}{\bibfnamefont{B.}~\bibnamefont{Hambrecht}},
  \bibinfo{author}{\bibfnamefont{P.}~\bibnamefont{Lugan}},
  \bibinfo{author}{\bibfnamefont{D.}~\bibnamefont{Cl{\'e}ment}},
  \bibinfo{author}{\bibfnamefont{L.}~\bibnamefont{Sanchez-Palencia}},
  \bibinfo{author}{\bibfnamefont{P.}~\bibnamefont{Bouyer}}, \bibnamefont{and}
  \bibinfo{author}{\bibfnamefont{A.}~\bibnamefont{Aspect}},
  \bibinfo{journal}{Nature (London)} \textbf{\bibinfo{volume}{453}},
  \bibinfo{pages}{891} (\bibinfo{year}{2008}).

\bibitem[{\citenamefont{Roati et~al.}(2008)\citenamefont{Roati, d'Errico,
  Fallani, Fattori, Fort, Zaccanti, Modugno, Modugno, and
  Inguscio}}]{Inguscio:AndersonBEC:N08}
\bibinfo{author}{\bibfnamefont{G.}~\bibnamefont{Roati}},
  \bibinfo{author}{\bibfnamefont{C.}~\bibnamefont{d'Errico}},
  \bibinfo{author}{\bibfnamefont{L.}~\bibnamefont{Fallani}},
  \bibinfo{author}{\bibfnamefont{M.}~\bibnamefont{Fattori}},
  \bibinfo{author}{\bibfnamefont{C.}~\bibnamefont{Fort}},
  \bibinfo{author}{\bibfnamefont{M.}~\bibnamefont{Zaccanti}},
  \bibinfo{author}{\bibfnamefont{G.}~\bibnamefont{Modugno}},
  \bibinfo{author}{\bibfnamefont{M.}~\bibnamefont{Modugno}}, \bibnamefont{and}
  \bibinfo{author}{\bibfnamefont{M.}~\bibnamefont{Inguscio}},
  \bibinfo{journal}{Nature (London)} \textbf{\bibinfo{volume}{453}},
  \bibinfo{pages}{895} (\bibinfo{year}{2008}).

\bibitem[{\citenamefont{Mackinnon}(1994)}]{MacKinnon:CriticalExp:JPC94}
\bibinfo{author}{\bibfnamefont{A.}~\bibnamefont{Mackinnon}},
  \bibinfo{journal}{J. Phys.: Condes. Matter} \textbf{\bibinfo{volume}{6}},
  \bibinfo{pages}{2511} (\bibinfo{year}{1994}).

\bibitem[{\citenamefont{Lewenstein et~al.}(2007)\citenamefont{Lewenstein,
  Sanpera, Ahufinger, Damski, Sen, and
  Sen}}]{Lewenstein:UltracoldSolidState:ADVP07}
\bibinfo{author}{\bibfnamefont{M.}~\bibnamefont{Lewenstein}},
  \bibinfo{author}{\bibfnamefont{A.}~\bibnamefont{Sanpera}},
  \bibinfo{author}{\bibfnamefont{V.}~\bibnamefont{Ahufinger}},
  \bibinfo{author}{\bibfnamefont{B.}~\bibnamefont{Damski}},
  \bibinfo{author}{\bibfnamefont{A.}~\bibnamefont{Sen}}, \bibnamefont{and}
  \bibinfo{author}{\bibfnamefont{U.}~\bibnamefont{Sen}},
  \bibinfo{journal}{Advances in Physics} \textbf{\bibinfo{volume}{56}},
  \bibinfo{pages}{243} (\bibinfo{year}{2007}).

\bibitem[{\citenamefont{Moore et~al.}(1995)\citenamefont{Moore, Robinson,
  Bharucha, Sundaram, and Raizen}}]{Raizen:QKRFirst:PRL95}
\bibinfo{author}{\bibfnamefont{F.~L.} \bibnamefont{Moore}},
  \bibinfo{author}{\bibfnamefont{J.~C.} \bibnamefont{Robinson}},
  \bibinfo{author}{\bibfnamefont{C.~F.} \bibnamefont{Bharucha}},
  \bibinfo{author}{\bibfnamefont{B.}~\bibnamefont{Sundaram}}, \bibnamefont{and}
  \bibinfo{author}{\bibfnamefont{M.~G.} \bibnamefont{Raizen}},
  \bibinfo{journal}{Phys. Rev. Lett.} \textbf{\bibinfo{volume}{75}},
  \bibinfo{pages}{4598} (\bibinfo{year}{1995}).

\bibitem[{\citenamefont{Casati et~al.}(1989)\citenamefont{Casati, Guarneri, and
  Shepelyansky}}]{Casati:IncommFreqsQKR:PRL89}
\bibinfo{author}{\bibfnamefont{G.}~\bibnamefont{Casati}},
  \bibinfo{author}{\bibfnamefont{I.}~\bibnamefont{Guarneri}}, \bibnamefont{and}
  \bibinfo{author}{\bibfnamefont{D.~L.} \bibnamefont{Shepelyansky}},
  \bibinfo{journal}{Phys. Rev. Lett.} \textbf{\bibinfo{volume}{62}},
  \bibinfo{pages}{345} (\bibinfo{year}{1989}).

\bibitem[{\citenamefont{Szriftgiser et~al.}(2003)\citenamefont{Szriftgiser,
  Lignier, Ringot, Garreau, and Delande}}]{AP:ChaosQTransp:CNSNS:2003}
\bibinfo{author}{\bibfnamefont{P.}~\bibnamefont{Szriftgiser}},
  \bibinfo{author}{\bibfnamefont{H.}~\bibnamefont{Lignier}},
  \bibinfo{author}{\bibfnamefont{J.}~\bibnamefont{Ringot}},
  \bibinfo{author}{\bibfnamefont{J.~C.} \bibnamefont{Garreau}},
  \bibnamefont{and} \bibinfo{author}{\bibfnamefont{D.}~\bibnamefont{Delande}},
  \bibinfo{journal}{Commun. Nonlin. Sci. Num. Simul.}
  \textbf{\bibinfo{volume}{8}}, \bibinfo{pages}{301} (\bibinfo{year}{2003}).

\bibitem[{\citenamefont{Ringot et~al.}(2001)\citenamefont{Ringot, Szriftgiser,
  and Garreau}}]{AP:RamanSpectro:PRA01}
\bibinfo{author}{\bibfnamefont{J.}~\bibnamefont{Ringot}},
  \bibinfo{author}{\bibfnamefont{P.}~\bibnamefont{Szriftgiser}},
  \bibnamefont{and} \bibinfo{author}{\bibfnamefont{J.~C.}
  \bibnamefont{Garreau}}, \bibinfo{journal}{Phys. Rev. A}
  \textbf{\bibinfo{volume}{65}}, \bibinfo{pages}{013403}
  (\bibinfo{year}{2001}).

\bibitem[{\citenamefont{Ringot et~al.}(1999)\citenamefont{Ringot, Lecoq,
  Garreau, and Szriftgiser}}]{AP:DiodeMod:EPJD99}
\bibinfo{author}{\bibfnamefont{J.}~\bibnamefont{Ringot}},
  \bibinfo{author}{\bibfnamefont{Y.}~\bibnamefont{Lecoq}},
  \bibinfo{author}{\bibfnamefont{J.~C.} \bibnamefont{Garreau}},
  \bibnamefont{and}
  \bibinfo{author}{\bibfnamefont{P.}~\bibnamefont{Szriftgiser}},
  \bibinfo{journal}{Eur. Phys. J. D} \textbf{\bibinfo{volume}{7}},
  \bibinfo{pages}{285} (\bibinfo{year}{1999}).

\bibitem[{not({\natexlab{a}})}]{note:chaotic}
\bibinfo{note}{{The standard kicked rotor becomes fully chaotic for $K > 5$,
  but our quasi-periodic version is chaotic for much smaller values $K > 2$.}}

\bibitem[{\citenamefont{Wegner}(1976)}]{Wegner:ScalingMobilityEdge:ZFP76}
\bibinfo{author}{\bibfnamefont{F.}~\bibnamefont{Wegner}}, \bibinfo{journal}{Z.
  Phys.} \textbf{\bibinfo{volume}{B25}}, \bibinfo{pages}{327}
  (\bibinfo{year}{1976}).

\bibitem[{not({\natexlab{b}})}]{note:Raman}
\bibinfo{note}{{The resolution of the Raman setup is chosen so that the width
  of this velocity class is 5 normalized momentum units. We experimentally
  verified that such a resolution produces a broadening of momentum
  distributions of less than 10\%.}}

\bibitem[{\citenamefont{Fisher and
  Barber}(1972)}]{FisherBarber:FiniteSizeScal:PRL72}
\bibinfo{author}{\bibfnamefont{M.~E.} \bibnamefont{Fisher}} \bibnamefont{and}
  \bibinfo{author}{\bibfnamefont{M.~N.} \bibnamefont{Barber}},
  \bibinfo{journal}{Phys. Rev. Lett.} \textbf{\bibinfo{volume}{28}},
  \bibinfo{pages}{1516} (\bibinfo{year}{1972}).

\bibitem[{\citenamefont{Slevin and
  Ohtsuki}(1999)}]{Slevin:ScalingAnderson:PRL99}
\bibinfo{author}{\bibfnamefont{K.}~\bibnamefont{Slevin}} \bibnamefont{and}
  \bibinfo{author}{\bibfnamefont{T.}~\bibnamefont{Ohtsuki}},
  \bibinfo{journal}{Phys. Rev. Lett.} \textbf{\bibinfo{volume}{82}},
  \bibinfo{pages}{382} (\bibinfo{year}{1999}).

\bibitem[{\citenamefont{Ohtsuki and
  Kawarabayashi}(1997)}]{Ohtsuki:AndersonTrans:JPSJ97}
\bibinfo{author}{\bibfnamefont{T.}~\bibnamefont{Ohtsuki}} \bibnamefont{and}
  \bibinfo{author}{\bibfnamefont{T.}~\bibnamefont{Kawarabayashi}},
  \bibinfo{journal}{J. Phys. Soc. Jpn.} \textbf{\bibinfo{volume}{66}},
  \bibinfo{pages}{314} (\bibinfo{year}{1997}).

\bibitem[{\citenamefont{Stauffer and Aharony}(1994)}]{Stauffer:Percolation:94}
\bibinfo{author}{\bibfnamefont{D.}~\bibnamefont{Stauffer}} \bibnamefont{and}
  \bibinfo{author}{\bibfnamefont{A.}~\bibnamefont{Aharony}},
  \textit{\bibinfo{title}{{Introduction to Percolation Theory}}}
  (\bibinfo{publisher}{{Taylor and Francis}}, \bibinfo{year}{1994}),
  \bibinfo{edition}{2nd} ed.

\bibitem[{not({\natexlab{c}})}]{note:CriticalExp}
\bibinfo{note}{{As is well known from numerical experiments on the Anderson
  model, the finite-size scaling method slightly underestimates the critical
  exponents \cite{MacKinnon:CriticalExp:JPC94}.}}

\end{thebibliography}

\end{document}